\documentclass[aps, prb, reprint]{revtex4-1}
\usepackage{lipsum}
\usepackage[version=3]{mhchem} 
\usepackage{miller}
\usepackage{graphicx}
\usepackage{amsmath}
\usepackage{color}
\usepackage[normalem]{ulem}
\usepackage{pdfpages}
\usepackage{etoolbox}
\makeatletter
\patchcmd{\@outputpage@head}{\@ifx{\LS@rot\@undefined}{}{\LS@rot}}{}{}{}
\makeatother

\begin{document}
\title{Band inversion driven by electronic correlations at the (111) \ce{LaAlO3}/\ce{SrTiO3} interface} 
\author{A.~M.~R.~V.~L.~Monteiro$^{1}$}
\email{A.M.Monteiro@tudelft.nl}
\author{M.~Vivek$^{2}$}
\author{D.~J.~Groenendijk$^{1}$}
\author{P.~Bruneel$^{2}$}
\author{I.~Leermakers$^{3}$}
\author{U.~Zeitler$^{3}$}
\author{M.~Gabay$^{2}$}
\author{A.~D.~Caviglia$^{1}$}
\affiliation{
$^{1}$Kavli Institute of Nanoscience, Delft University of Technology, P.O. Box 5046, 2600 GA Delft, The Netherlands.\\
$^{2}$Laboratoire de Physique des Solides, Universit\'{e} Paris-Sud 11, Universit\'{e} Paris Saclay, CNRS UMR 8502, 91405 Orsay Cedex, France.\\
$^{3}$High Field Magnet Laboratory (HFML-EFML), Radboud University Nijmegen, 6525 ED Nijmegen, The Netherlands.}
%\date{\today}

\begin{abstract}
Quantum confinement at complex oxide interfaces establishes an intricate hierarchy of the strongly correlated $d$-orbitals which is widely recognized as a source of emergent physics.
The most prominent example is the (001) \ce{LaAlO3}/\ce{SrTiO3} (LAO/STO) interface, which features a dome-shaped phase diagram of superconducting critical temperature and spin--orbit coupling (SOC) as a function of electrostatic doping, arising from a selective occupancy of $t_{2g}$ orbitals of different character. Here we study (111)-oriented LAO/STO interfaces---where the three $t_{2g}$ orbitals contribute equally to the sub-band states caused by confinement---and investigate the impact of this unique feature on electronic transport. We show that transport occurs through two sets of electron-like sub-bands, and the carrier density of one of the sets shows a non-monotonic dependence on the sample conductance. Using tight-binding modeling, we demonstrate that this behavior stems from a band inversion driven by on-site Coulomb interactions. The balanced contribution of all $t_{2g}$ orbitals to electronic transport is shown to result in strong SOC with reduced electrostatic modulation. 
\end{abstract}
% insert suggested PACS numbers in braces on next line
\maketitle %must follow title, authors, abstract and \pacs

%We uncover an unexpected transition from one- to two-carrier transport as a function of electrostatic doping\textcolor{red}{, finding that a second electron band is populated at the cost of the first.}

%%%%%%%%%%%%%%%%%%%%%%%%%%%%%%%%%%%%%%%%%%%%%%%%%%%%%%%%%%%%%%%%%%%%%%%%%%%%%%%%%%%%%%%%%%%%%%%%%%%%

Complex oxide interfaces display a variety of emergent physical properties that arise from their highly correlated $d$-electrons and are therefore absent in conventional semiconductor quantum wells\,\cite{dagotto2005complexity,hwang2012emergent}. The two-dimensional electron system (2DES) at the interface between \ce{LaAlO3} (LAO) and (001)-oriented \ce{SrTiO3} (STO) is the prototypical oxide quantum well\,\cite{ohtomo2004high}, featuring several interesting phenomena that include 2D superconductivity\,\cite{reyren2007superconducting} and Rashba spin--orbit coupling (SOC)\,\cite{caviglia2010tunable,shalom2010tuning}. The hierarchy of $d$-orbitals with different symmetries imposed by two-dimensional confinement has been recognized as a key element in determining the properties of the system\,\cite{salluzzo2009orbital}. In particular, it has been proposed that the dome-shaped behavior of the superconducting critical temperature ($T_{\mathrm{c}}$) and SOC strength with electrostatic doping is related to the selective occupancy of orbitals of different character, detected by a transition from one to two-carrier transport\,\cite{joshua2011universal}. On the other hand, recent works have shown that the crystallographic direction of confinement is a powerful tool enabling selective modification of this band hierarchy\,\cite{herranz2012high,herranz2015engineering,pesquera2014two}. (111)-oriented LAO/STO interfaces are of particular interest, since the sub-band structure due to quantum confinement preserves the $t_{2g}$ manifold symmetry along this direction~\cite{doennig2013massive,walker2014control,song2018direct}. Transport studies have shown that the system condenses into a superconducting ground state~\cite{monteiro2017two, davis2017magnetoresistance, rout2017link} and proposed a link between $T_{\mathrm{c}}$ and SOC ~\cite{rout2017link}. More strikingly, field-effect measurements have brought to light an unconventional behavior of the Hall coefficient ($R_\textrm{H}$), which has been interpreted as a signature of a hole-like band~\cite{davis2016electrical, davis2017anisotropic, rout2017link}.

\begin{figure}[ht!]
\includegraphics[width=\linewidth]{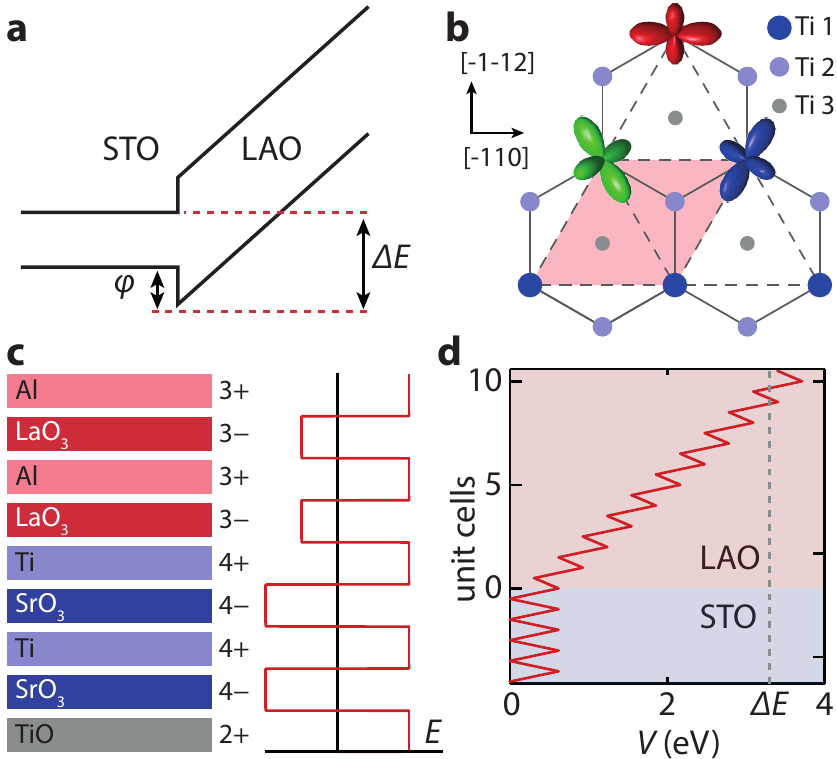}%
\caption{
(a) Band diagram of the LAO/STO interface before electronic reconstruction. $\Delta E$: critical potential build-up. $\phi$: valence-band offset.
(b) Top view of three consecutive (111) \ce{Ti^4+} layers. The red shaded area represents the unit cell cross-section of a bilayer. The three $t_{2g}$ orbitals are shown to evidence their equivalent projection onto the 2DES plane.
(c) Left: stacking of ionic planes across the interface. The bottom-most \ce{Ti^4+} plane is considered to react with oxygen to form \ce{TiO^2+}. Right: resulting electric field across the interface before the electronic reconstruction takes place.
(d) Electrostatic potential as a function of the number of unit cells.
\label{fig:pol_cat}}%
\end{figure}

In this work, we investigate the electronic properties of (111)-oriented LAO/STO interfaces and show that (i) transport occurs solely through electron-like sub-bands and (ii) a sub-band inversion triggered by local Coulomb interactions is key to explain the unusual behavior of $R_\textrm{H}$. Importantly, we show that this inversion occurs between two sets of $t_{2g}$ sub-bands, each with a balanced contribution of $d_{xy}$, $d_{yz}$ and $d_{xz}$ character. As a direct consequence of this unique feature, SOC is strong and displays reduced electrostatic tunability.

Initially, the study of LAO/STO interfaces was restricted to the (001) crystallographic direction, where the emergence of conduction was originally explained in terms of the polar-catastrophe scenario\,\cite{ohtomo2004high,nakagawa2006why}. In this model, a polar discontinuity arises at the interface between LAO and (001) STO\,\cite{ohtomo2004high} as a consequence of the stacking of charged ionic LAO planes (with alternating valency of $+1e$ and $-1e$) over the neutral STO planes. As a result, the voltage grows with the thickness of the LAO film until the built-in potential becomes larger than $\Delta E$ (Fig.\,\ref{fig:pol_cat}a). At a critical thickness $t_c \approx 3.5\,\mathrm{u.c.}$, this triggers an electronic reconstruction in which half an electron per unit cell is transferred from the surface of the LAO film to Ti $3d$ states at the interface\,\cite{thiel2006tunable,reinle2012tunable}. More recent works have shown that the polar field triggers the spontaneous formation of surface oxygen vacancies, leading to interface conductivity\,\cite{bristowe2011surface,Yu2014}.

\begin{figure}[ht!]
\includegraphics[width=\linewidth]{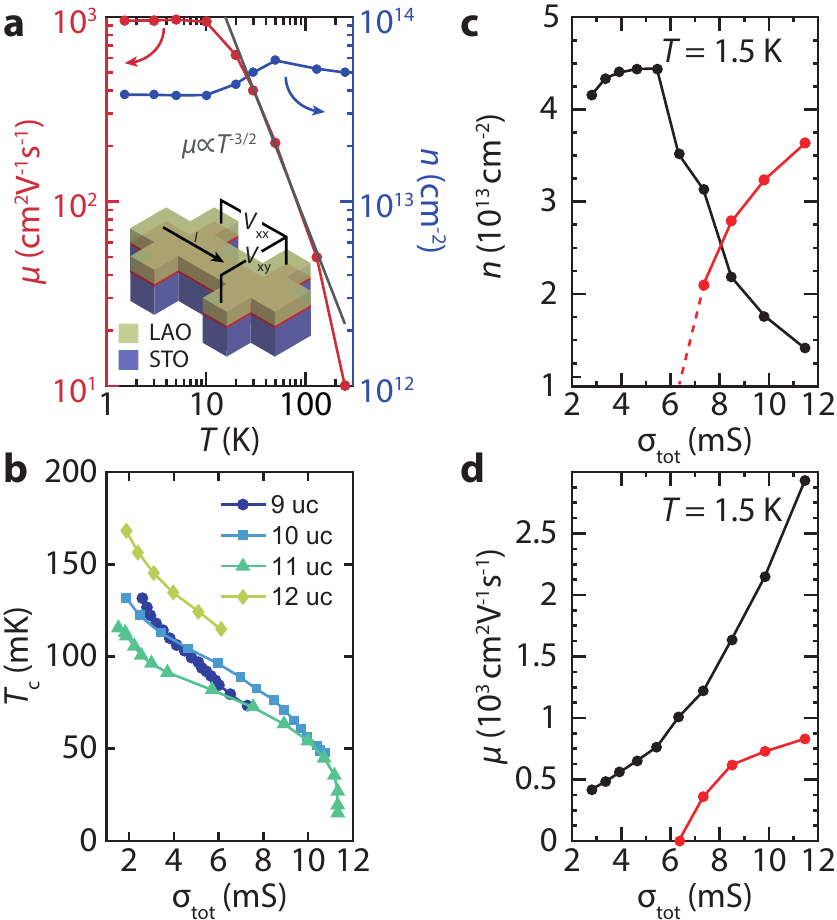}%
\caption{
(a) Carrier density ($n$) and mobility ($\mu$) as a function of temperature ($T$) measured for the pristine state. Inset: schematic representation of the measurement configuration. (b) Superconducting critical temperature ($T_{\mathrm{c}}$) as a function of sample conductance ($\sigma_{\mathrm{tot}}$) for different thicknesses of the LAO film. (c) Carrier densities and (d) mobilities as a function of $\sigma_{\mathrm{tot}}$.
\label{fig:n_mu}}%
\end{figure}

Having proposed a possible solution for the polar instability at (111)-oriented LAO/STO interfaces, we investigate the evolution of electronic properties as a function of temperature and electrostatic doping. The temperature dependence of carrier density ($n$) and mobility ($\mu$) for a 9 u.c.~LAO/STO (111) interface is shown in Fig.\,\ref{fig:n_mu}a. In the pristine state, the Hall effect remains linear down to $1.5\,\mathrm{K}$ in a range of $10\,\mathrm{T}$. The extracted carrier density remains fairly constant around $3$-$5 \times 10^{13}\,\mathrm{cm^{-2}}$ in the entire temperature range. The mobility increases rapidly from $10\,\mathrm{cm^{2}V^{-1}s^{-1}}$ at room temperature to a maximum value of $1000\,\mathrm{cm^{2}V^{-1}s^{-1}}$ at $1.5\,\mathrm{K}$, with saturation occurring below $10\,\mathrm{K}$. The gray line represents the phonon-limited mobility $\mu_{\mathrm{ph}} \propto T^{-3/2}$, showing good accordance with the data at high temperatures. Moreover, the carrier density values obtained are comparable with those reported for (001)-oriented interfaces. 

At $1.5\,\textrm{K}$, we use a back-gate geometry to perform high-field magnetotransport measurements as a function of electrostatic doping. At high conductance values, a transition from linear to non-linear Hall curves is observed, indicating a transition from one to two-carrier transport. In contrast with previous works~\cite{davis2016electrical, davis2017anisotropic, rout2017link}, the observation of this non-linearity enables us to unequivocally ascertain that the two bands involved in transport are electron-like, and in the Supplemental Material we analytically show that the evolution of $R_\textrm{H}$ as a function of $B$ is incompatible with an electron-hole scenario. Figures\,\ref{fig:n_mu}c and d show the extracted values of carrier density ($n_{1,2}$) and mobility ($\mu_{1,2}$) by fitting the Hall curves to a two-band model (see Supplemental Material). The appearance of the second band at $\sigma \approx 6\,\mathrm{mS}$ is readily evident: at this point, $n_2$ increases rapidly, seemingly at the expense of $n_1$. Moreover, the second band has a mobility which is roughly 3 times smaller than the first band. 

In the millikelvin regime, the system condenses into a superconducting ground state\cite{monteiro2017two} and measurements in the same conductance range reveal a monotonic decrease of $T_{\mathrm{c}}$. This suggests that superconductivity is unaffected by the population of the second electron sub-band at the expense of the first. As shown in Fig.\,\ref{fig:n_mu}b, this behavior is consistently observed in several samples, with LAO thicknesses ranging from 9 to $12\,\mathrm{u.c.}$. This is in stark contrast with (001)-oriented interfaces, where the maximum of the superconducting dome occurs concomitantly with the onset of population of the $d_{xz,yz}$ bands at the Lifshitz point. In the (111) crystallographic direction, all the $t_{2g}$ orbitals have the same geometrical projection onto the 2DES plane (see Fig.\,\ref{fig:pol_cat}b), therefore the observed transition must have an intrinsically different origin than the one observed in the (001) counterpart.

%This is in stark contrast with \hkl(001)-oriented interfaces, where it is well established that a Lifshitz transition takes place between $d_{xy}$ and $d_{xz,yz}$ orbital populations at a well defined electrostatic doping level.} Interestingly, the maximum of the superconducting dome occurs concomitantly with the onset of population of the $d_{xz,yz}$ bands at the Lifshitz point.

\begin{figure}[ht!]
\includegraphics[width=\linewidth]{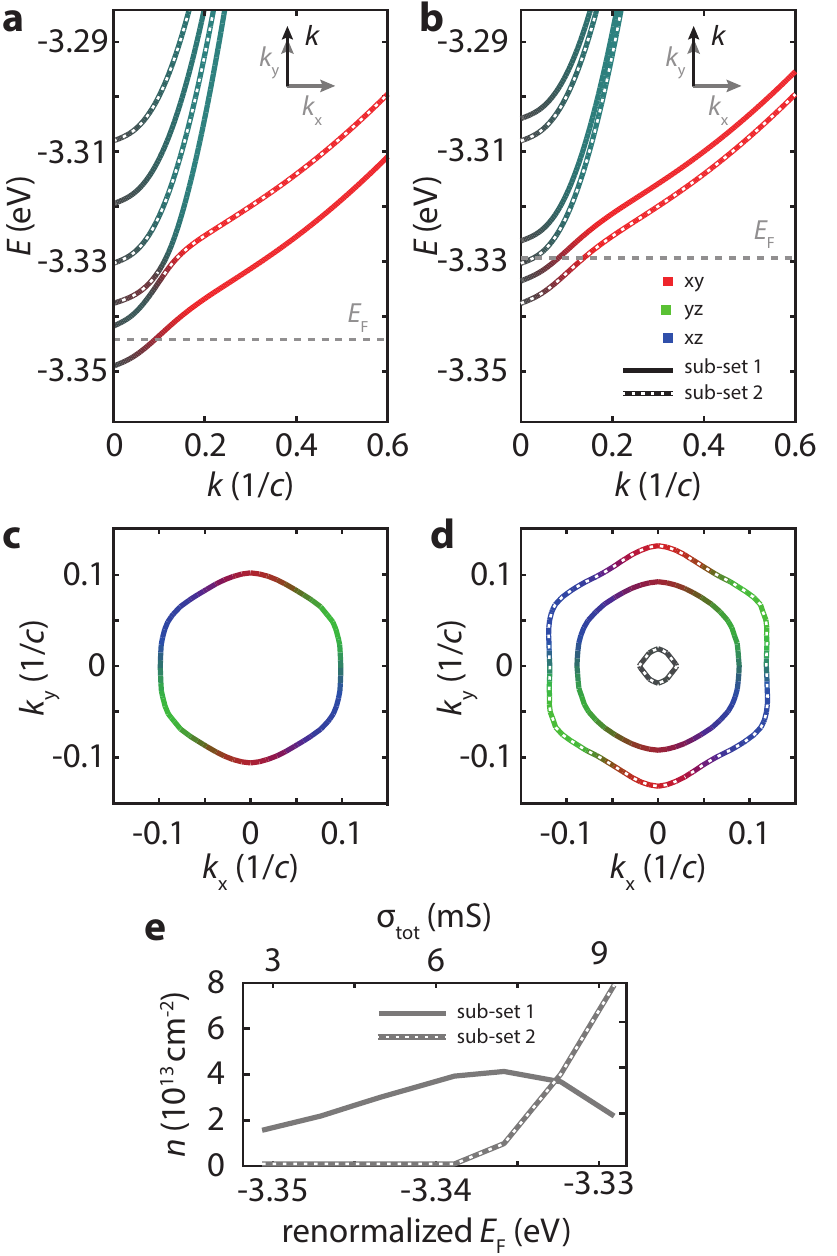}
\caption{(a,b) Band structure at low and high filling, respectively. Dashed gray line indicates the renormalized Fermi level. Color indicates the orbital character. Stoke indicates the band sub-set. Inset denotes the direction of the $k$-vector in the reciprocal space. 
(c,d) Corresponding Fermi surfaces. (e) Evolution of the carrier density pertaining to the first ($n_1$) and second ($n_2$) sub-set of bands as a function of renormalized Fermi level and respective sample conductance.
\label{fig:SC}}%
\end{figure}

%Tight-binding calculations of the electronic structure of the (111)STO surface derived from photoemission spectra \cite{walker2014control, rodel2014orientational} show that 
The sub-band structure was determined from Poisson-Schroedinger calculations and dispersions consequently derived by means of tight-binding modeling (Supplemental Material). For the experimentally accessible range of carrier concentrations, two sets of sub-bands lie close to the Fermi energy $E_{\mathrm{F}}$.
These two sets of sub-bands, labeled 1 and 2 in Fig. \ref{fig:SC}, each contain six branches. However, due to time reversal symmetry, there are only 3 different energies per set, thus leading to a six-band low energy model. In our tight-binding calculations we include the effects of (i) confinement, (ii) bulk SOC, (iii) trigonal field, and (iv) Hubbard type on-site interactions between like (U) and unlike (U$^{'}$) orbitals. Coulomb terms cause the bands to shift by unequal amounts resulting in bands crossings and in changes in the individual carrier concentrations of the bands. In order to keep the total carrier density constant before and after the inclusion of interactions, the Fermi level renormalises. This renormalisation of the Fermi level is performed in a self-consistent way (see Supplemental Material for further details on the theoretical model).
The resulting band structures are plotted in Fig.\,\ref{fig:SC}a-b where we show  the Energy vs. momentum ($E$ vs.~$k$) along the $k_x=0$ direction for two different filling factors. $k_y$  corresponds to $\Gamma$M of the hexagonal Brillouin Zone (BZ) and $k_x$ corresponds to $\Gamma$K of the hexagonal BZ. Both $k_x$  and $k_y$  are in units of $1/c$, where $c=\sqrt{2/3}a$, and $a$ is the Ti-Ti inter-atomic distance.
Careful analysis of Fig.\,\ref{fig:SC}a-b readily highlights the crucial role of electron correlations in reproducing our experimental observations. At low $E_\textrm{F}$ (Fig.\,\ref{fig:SC}a), only the first set of sub-bands is populated. At high $E_\textrm{F}$ (Fig.\,\ref{fig:SC}b), the second set of sub-bands, which extends deeper into the substrate, becomes populated and---most importantly---a band inversion takes place. The second set of sub-bands becomes lower in energy, while the first sub-set is pushed upwards. The consequences of this can be more clearly seen in the corresponding Fermi surfaces plotted in Fig.\,\ref{fig:SC}c and d, where the contour of the first set of sub-bands is reduced with increasing $E_\textrm{F}$. Conversely, it is evident in Fig.\,\ref{fig:SC}d that the second sub-band becomes heavily populated, its contour becoming larger than that of the first sub-band.
It is worth underscoring that, while the orbital character of each band is highly dependent on the crystallographic direction in the BZ, their overall contributions to electronic transport are nearly equal. The concentrations of the carriers in each band are summed for each sub-set and are shown in Fig.\,\ref{fig:SC}e as a function of $E_{F}$.  The resemblance with the experimental data is striking: at low filling only the first set of sub-bands contributes to transport and, at a critical filling, the population of the second set of sub-bands starts increasing, concomitantly with a decline of the population of the first one. Our model highlights that, in contrast with the (001) case, the transition from one to two-carrier transport in the (111) direction stems from the occupation of a second set of $t_{2g}$ sub-bands as a consequence of Coulomb repulsion.  

\begin{figure}[ht!]
\includegraphics[width=\linewidth]{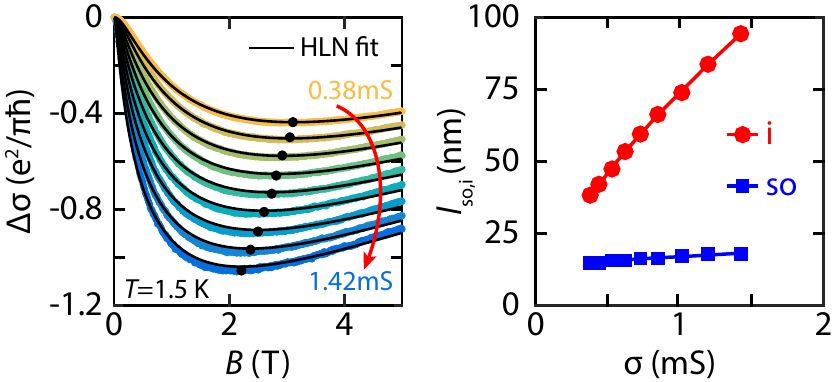}%
\caption{(a) Variation of conductance $\Delta \sigma$ as a function of $B$-field for different levels of electrostatic doping. Black dots: $B_{\mathrm{min}}$. Black lines: fit to the HLN equation. (b) Extracted characteristic lengths $l_{\mathrm{i,so}}$.
\label{fig:HLN}}%
\end{figure}

To investigate the effects of the orbital hierarchy of (111)-oriented LAO/STO on SOC, we analysed the field dependence of the magnetoconductance (MC) as a function of electrostatic doping. We restrict our analysis to low conductance values, where the Hall effect is linear and the classical magnetoconductance contribution is negligible (see Supplemental Material). As shown in Fig.\,\ref{fig:HLN}a, negative MC is observed in the entire range of conductance explored, in accordance with previous work\,\cite{rout2017link}.
For a 2D diffusive metallic system placed in a perpendicular magnetic field ($B$), the quantum corrections to conductance are given by the Hikami-Larkin-Nagaoka (HLN) model\,\cite{hikami1980spin}:

\begin{equation}
\begin{split}
\frac{\Delta \sigma (B)}{\sigma_0} =& - \frac{1}{2} \Psi \left( \frac{1}{2} + \frac{B_\mathrm{i}}{B} \right) - \frac{1}{2} \ln \frac{B_{\mathrm{i}}}{B} \\ & - \Psi \left( \frac{1}{2} + \frac{B_\mathrm{i} + B_\mathrm{SO} }{B} \right) + \ln \frac{B_{\mathrm{i}} + B_{\mathrm{SO}}}{B} \\ & - \frac{1}{2} \Psi \left( \frac{1}{2} + \frac{B_\mathrm{i} + 2B_\mathrm{SO} }{B} \right) + \frac{1}{2} \ln \frac{B_{\mathrm{i}} + 2B_{\mathrm{SO}}}{B},
\end{split}
\end{equation}

where $\Psi$ is the digamma function and $B_{\mathrm{i,SO}}$ are the effective fields related to the inelastic and spin--orbit relaxation lengths, respectively. Fig.\,\ref{fig:HLN}a shows $\Delta\sigma$ and the respective quantum correction from the HLN model (black lines). $\Delta\sigma$ displays a local minimum at a field $B_{\mathrm{min}}$, which indicates the point where weak antilocalization (WAL) is overcome by weak localization (WL). It has been demonstrated that $B_{\mathrm{min}}$ is proportional to the characteristic magnetic field $B_{\mathrm{SO}}$\,\cite{liang2015nonmonotonically}. Therefore, the gradual shift of $B_{\mathrm{min}}$ to smaller values of $B$ as a function of electrostatic doping indicates a monotonic decrease of the SOC strength. In contrast with previous work~\cite{rout2017link}, no classical component was required to fit the data and the local minima of the data are well captured. 
The characteristic lengths of inelastic and spin--orbit scattering $l_{\mathrm{i,so}}$ are related to the effective fields by $B_{\mathrm{i,SO}} = \hbar / 4el_{\mathrm{i,SO}}^2$ and are shown as a function of applied electrostatic doping in Fig.\,\ref{fig:HLN}b. We find that $l_{\mathrm{SO}} < l_{\mathrm{i}}$ throughout the entire range, indicating WAL. Moreover, we observe a relatively small value of $l_{\mathrm{SO}}$ which exhibits a limited variation with electrostatic doping, indicating that SO interactions are strong ($\varepsilon_\textrm{SO}\approx 4.26\,\textrm{meV}$ at $0.38\,\textrm{mS}$), but overall display reduced tunability with respect to the (001)-oriented case. This can be understood by recalling the hallmark feature of this crystallographic direction---the identical projection of all $t_{2g}$ orbitals onto the 2DES plane. 
The magnitude of $B_{\mathrm{SO}}$, which is proportional to the out-of-plane component of the orbitals involved in transport\,\cite{khalsa2013theory}, is therefore expected to be large and independent of band occupation, in very good agreement with our experimental results.

%eps_SO from H_parallel: 2.78 meV

In summary, we have studied (111)-oriented LAO/STO interfaces where $t_{2g}$ manifold splitting by quantum confinement is absent. We demonstrate that transport occurs through electron-like sub-bands and on-site correlations drive an 
inversion between two sets of $t_{2g}$ sub-bands, each containing a balanced contribution of all three orbital characters. This captures the non-monotonic dependence of $R_\textrm{H}$ on electrostatic doping and rules out the presence of a hole-like band. The results of this work strongly underline the importance of orbital hierarchy and electron-electron interactions in determining the properties of LAO/STO interfaces.

%%%%%%%%%%%%%%%%%%%%%%%%%%%%%%%%%%%%%%%%%%%%%%%%%%%%%%%%%%%%%%%%%%%%%%%%%%%%%%%%%%%%%%%%%%%%%%%%%%%%
\begin{acknowledgments}
This work was supported by The Netherlands Organisation for Scientific Research (NWO/OCW) as part of the Frontiers of Nanoscience program (NanoFront) and the DESCO program, by the Dutch Foundation for Fundamental Research on Matter (FOM). The research leading to these results has received funding from the European Research Council under the European Union's H2020 programme/ ERC Grant Agreement No.~677458 and project Quantox of QuantERA ERA-NET Cofund in Quantum Technologies.
Support from the French National Research Agency (ANR), project LACUNES No.~ANR-13-BS04-0006-01 is gratefully acknowledged.
\end{acknowledgments}

\bibliographystyle{apsrev4-1}
\bibliography{references}
\clearpage


\section*{Supplementary material}

# Supplemental Material for:
# Band inversion driven by electronic correlations at the (111) LaAlO$_3$/SrTiO$_3$ interface

A. M. R. V. L. Monteiro,[1, *] M. Vivek,[2] D. J. Groenendijk,[1]
P. Bruneel,[2] I. Leermakers,[3] U. Zeitler,[3] M. Gabay,[2] and A. D. Caviglia[1]

[1] Kavli Institute of Nanoscience, Delft University of Technology,
P.O. Box 5046, 2600 GA Delft, The Netherlands.
[2] Laboratoire de Physique des Solides, Université Paris-Sud 11,
Université Paris Saclay, CNRS UMR 8502, 91405 Orsay Cedex, France.
[3] High Field Magnet Laboratory (HFML-EFML), Radboud University Nijmegen, 6525 ED Nijmegen, The Netherlands.

## I. THEORETICAL MODELING

The (111)-oriented interface features a very different geometry than that of its (0 0 1)-oriented counterpart. When the cubic structure with Titanium (Ti) atoms is projected on one plane, three interspersing Ti lattices are found (Ti 1, 2 and 3), as shown in Fig. S1, which forms a hexagonal structure. In our modeling, only two layers are considered, assuming the coupling of the third layer to be weak as it is further away from the surface.

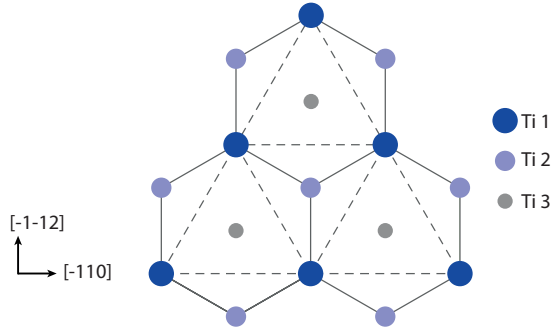

FIG. S1. Top view of three consecutive Ti$^{4+}$ layers in (111)-oriented STO. The number indicates the distance of the respective atomic layer to the STO surface. For theoretical calculations, only two Ti$^{4+}$ layers are taken in to account.

In a tight-binding formalism, there exist two kinds of hopping for the two layers-inter-layer and intra-layer, the amplitude of which depends on the distance between the concerned atoms. If only intra-layer coupling is added, three bands, one of each $t_{2g}$ orbitals, is observed, 4 times degenerate, twice in spin and twice for each layer. The intra-layer coupling is described by

$$a_{d_{xy}} = -2t_1 \cos(\sqrt{3} k_x) \tag{1}$$

$$a_{d_{yz}} = -2t_1 \cos(-\tfrac{\sqrt{3}}{2} k_x + \tfrac{3}{2} k_y) \tag{2}$$

$$a_{d_{xz}} = -2t_1 \cos(\tfrac{\sqrt{3}}{2} k_x + \tfrac{3}{2} k_y) \tag{3}$$

where $t_1 = 0.05\,\mathrm{eV}$ is the intra-layer hopping amplitude. The inter-layer coupling then splits this manifold into three bonding and three anti-bonding bands, twice degenerate in spin, each. The inter-layer coupling is described by

$$a'_{d_{xy}} = -t_2 - 2t_3 \cos(\tfrac{\sqrt{3}}{2} k_x) e^{-\mathrm{I} \tfrac{3}{2} k_y} \tag{4}$$

$$a'_{d_{yz}} = -t_3 (1 + e^{\mathrm{I}(\tfrac{\sqrt{3}}{2} k_x - \tfrac{3}{2} k_y)}) - t_2 e^{-\mathrm{I}(\tfrac{\sqrt{3}}{2} k_x + \tfrac{3}{2} k_y)} \tag{5}$$

$$a'_{d_{xz}} = -t_3 (1 + e^{-\mathrm{I}(\tfrac{\sqrt{3}}{2} k_x - \tfrac{3}{2} k_y)}) - t_2 e^{\mathrm{I}(\tfrac{\sqrt{3}}{2} k_x - \tfrac{3}{2} k_y)} \tag{6}$$

---

* A.M.Monteiro@tudelft.nl

where $t_3 = 1.6\,\mathrm{eV}$ and $t_2 = 0.07\,\mathrm{eV}$ are first and third nearest neighbour hoppings between layers. Since the 2DES resides on the STO side of the LAO/STO interface, one expects similar tight binding dispersions for (111)-oriented STO and LAO/STO, allowing us to use photoemission data for the (111) STO surface [1, 2] (this is indeed the case for the (001) orientation [3]). Only the lower energy bonding states are then required which lie close to the Fermi level. An inter-layer electrostatic potential is added to distinguish between the two layers in terms of their distance from the surface, which increases the distance between the bonding and the anti-bonding orbitals without changing their character. Using Dirac's notation, the Hamiltonian $H_o$ in basis of $\{dxy_{1,\sigma}, dyz_{1,\sigma}, dxz_{1,\sigma}, dxy_{2,\sigma}, dyz_{2,\sigma}, dxz_{2,\sigma}\}$ for the two layers and where 1, 2 is the layer index and $\sigma$ for spin $\uparrow, \downarrow$, is given by

$$H_{o;\sigma} = \sum_{P=(dxy,dyz,dxz)} |d_{P,1,\sigma}, d_{P,2,\sigma}\rangle \, \mathbf{M} \, \langle d_{P,2,\sigma}, d_{P,1,\sigma}| \qquad (7)$$

where $M = \begin{bmatrix} a_P & a'_P \\ a'^{*}_P & b_P \end{bmatrix}$ and $b_P = a_p + \Delta V$, $\Delta V$ being the difference of potential between the two layers owing to their different location with respect to the surface.

The next term in the modeling is the trigonal crystal field, the magnitude of which is given by $tf = 10\,\mathrm{meV}$, which arises as a result of the changed (111) symmetry at the surface. This lifts further the degeneracy at $\Gamma$ and is represented in the $\{d_{xy}, d_{yz}, d_{xz}\}$ basis by $H_{trf} = \begin{bmatrix} 0 & tf & tf \\ tf & 0 & tf \\ tf & tf & 0 \end{bmatrix}$. The trigonal field needs to be added separately to each layer and each spin. Bulk spin-orbit interaction, $H_{so} = \lambda \vec{L}.\vec{S}$, couples the orbital and spin degrees of freedom within each layer and changes the character of a band from pure $t_{2g}$ to a mixture of all three $t_{2g}$ orbitals. It also creates a spin-split off band at $\Gamma$.

The final term before the addition of correlations is the modeling of the quantum confinement at the surface of the (111)-oriented interface. Poisson-Schrodinger (P-S) calculations show that the bands introduced above give rise to sub-bands in the quantum well, created by the surface potential, and lying very close to each other. At least, two of these sets of sub-bands lie close to the Fermi level and must be considered in the effective tight-binding modeling. The Hamiltonian of one set of sub-bands can be constructed with $H_{0,\uparrow}, H_{0,\uparrow}, H_{1,so}, H_{2,so}$ and $H_{trf}$ in the basis of $|d_{P,1,\uparrow}, d_{P,2,\uparrow}, d_{P,1,\downarrow}, d_{P,2,\downarrow}\rangle$. A similar matrix exists for the sub-band set two with a shift of the bands by 20 meV as given by P-S calculations. Correlations become important in this modeling as the experimental data shows that the back-gate voltage tunes the number of carriers in the system and they increase with increasing back-gate voltage.

A Hartree-Fock interaction is then added which acts on individual orbitals. From therein stem two contributions to the correlation term, those between like (U) and unlike (U$'$) orbitals, whose product with the band populations of like and unlike orbitals respectively. The band populations take the following form for each orbital $\alpha \in \{xy, yz, xz\}$

$$< N_\alpha > = \sum_i < n_{i\alpha} > \qquad (8)$$

In Eq. 8, $< n_{i\alpha} >$ are the populations of the individual orbitals, $i$ is the band index and the summation is performed for bands with energies less than $E_F$ over all degrees of freedom; $< N_\alpha >$ is the total population of the $\alpha$ orbital. The addition of U and U$'$ gives rise to the following Hartree-Fock contribution

$$\mu_\alpha = U < N_\alpha > + U' \sum_{\gamma \neq \alpha} < N_\gamma > \qquad (9)$$

In Eq. 9 $\mu_\alpha$ plays the role of an orbital dependent chemical potential which shifts the bands. As $U$ and $U'$ are added, all orbitals are not equally punished which results in bands being shifted unequally causing band crossings. Before the addition of U and U$'$, the total population, $N_0$, is calculated and this should remain constant even after the addition of U and U$'$. At each iteration of the code, U and U$'$ shift the bands while the Fermi level remains unchanged. The total band population, $N_i$, is then calculated by summing up individual populations and compared to $N_0$. The Fermi level is re-adjusted to regain the original total population which again changes the band populations and causes the U and U$'$ to shift the bands at the next iteration. This process is repeated until self-consistency is achieved. Figure S2 shows the evolution of the band structure as one goes around the BZ. The filling chosen is such that, after correlations are added, both sub-bands are occupied. The bands are color-labeled according to their orbital character as in the main text. Upon changing the orientation along which one views the band structure, one finds that the lower most occupied band changes character. The integral over the entire BZ yields almost equal population of each orbital character within the manifold of a single sub-band and hence the average of the populations is considered in the calculations for Fig. 3e of the main text.

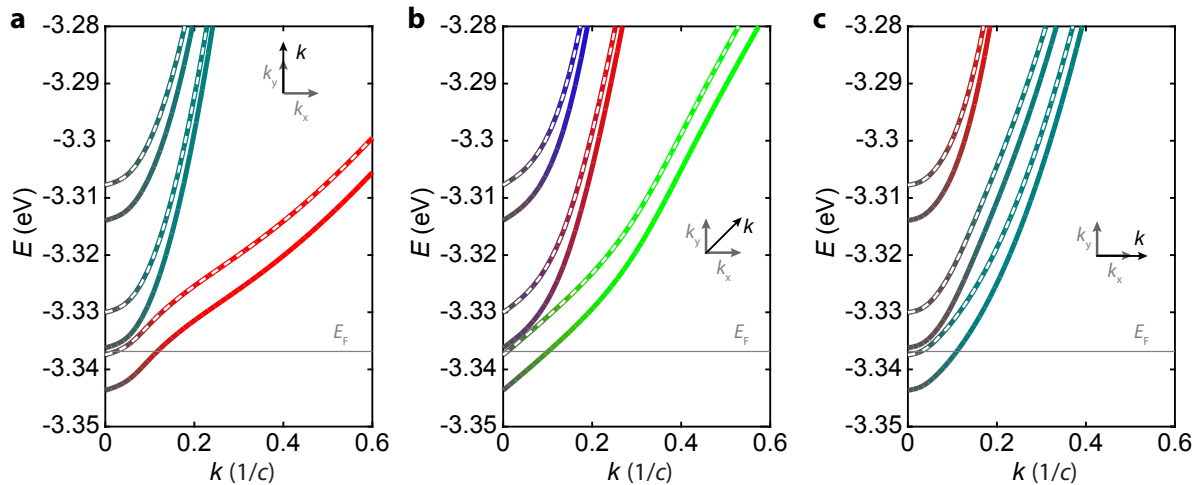

FIG. S2. Band structure decomposed into the three $t_{2g}$ orbitals of each sub-band in different orientations of the BZ for a filling of the bands where both sub-bands are occupied. Gray line indicates the renormalized Fermi level. (a) With $k$ along the $k_x = 0$ direction. (b) With $k$ along the $k_x = k_y$ direction. (c) With $k$ along the $k_y = 0$ direction.

## II. ANALYSIS OF THE HALL EFFECT AND MAGNETORESISTANCE DATA

The Hall data has been fitted to a two-carrier model given by the expression:

$$\rho_{xy} = \frac{(\sigma_1\mu_1 + \sigma_2\mu_2) + (\sigma_1\mu_2 + \sigma_2\mu_1)\mu_1\mu_2 B^2}{(\sigma_1 + \sigma_2)^2 + (\sigma_1\mu_2 + \sigma_1\mu_1)^2 B^2} B, \tag{10}$$

where $\sigma_{1,2}$ are the conductivities and $\mu_{1,2}$ are the mobilities of each one of the bands. The condition $\sigma_{\text{total}} = \sigma_1 + \sigma_2$ is imposed in the fit, since $\sigma_{\text{total}}$ is experimentally known. The experimental data and respective fits are shown in Fig. S3.

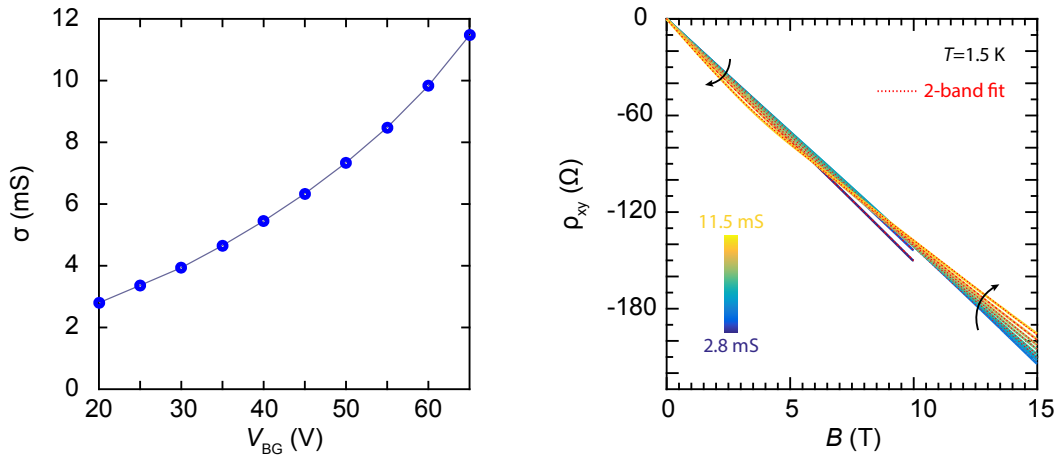

FIG. S3. Left: Total conductance as a function of applied back-gate voltage. Right: Hall resistance ($\rho_{xy}$) as a function of applied magnetic field ($B$) for different levels of electrostatic doping, measured at 1.5 K. Red dashed lines are the fits to a two-carrier model. Black arrows illustrate the trend of the slope at low and high magnetic fields.

The depopulation of the first band can also be qualitatively observed in the Hall traces, where the slope at low field follows the opposite trend of the slope at high field as a function electrostatic doping (denoted by the black arrows in Fig. S3). The parameters extracted from the fitting are shown in Fig. 2 of the main text.

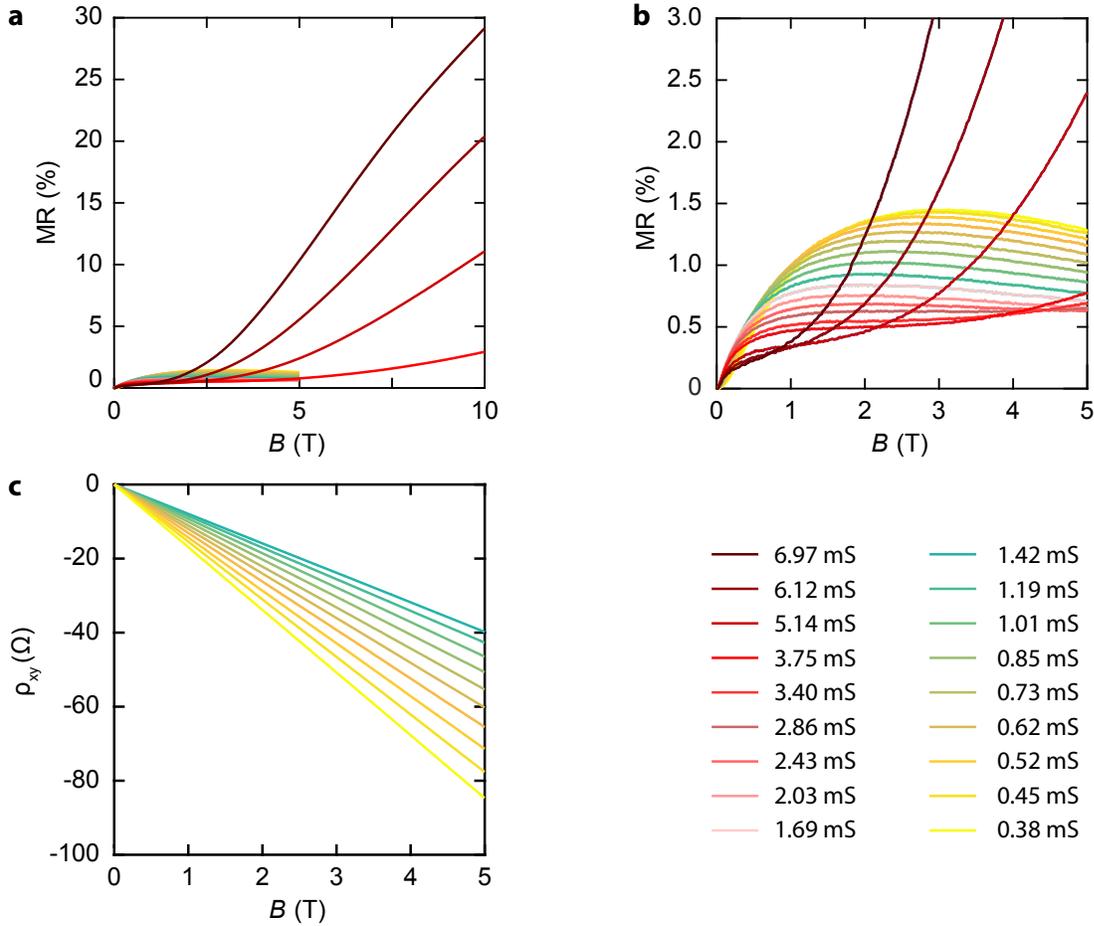

FIG. S4. Magnetoresistance as a function of applied of applied magnetic field for different levels of electrostatic doping. (a) Entire $B$-field range. (b) Zoom-in. Red-shaded curves correspond to curves which were not fitted in the MC analysis Fig. 4 of the main text. (c) Hall effect data corresponding to the same levels of electrostatic doping of the MC data of Fig. 4 of the main text.

Figure S4 shows the magnetoresistance (MR) as a function of applied magnetic field for different levels of electrostatic doping, where magnetoresistance is defined as

$$\text{MR} = \frac{\rho(B) - \rho(B=0)}{\rho(B=0)}. \tag{11}$$

The red traces indicate the electrostatic doping levels for which the Hall effect becomes non-linear, clearly indicting a two-carrier transport. As mentioned in the main text, these traces display a large classical $B^2$ component, which increases with electrostatic doping. This classical component can be as large as 30% at $B = 10\,\text{T}$ for the highest doping level.

## III. CARRIER TYPE OF THE 2 BANDS

In the two-band model, the Hall effect saturates towards a linear behaviour at sufficiently high magnetic fields and the sign of $R_\text{H}$ provides direct information on the carrier type of the higher-density band. In the case of an electron-like and a hole-like band, the high-field ($B \to \infty$) and low-field ($B \to 0$) limits of $R_\text{H}$ are given by

$$\lim_{B \to \infty} R_\text{H} = -\frac{1}{n_e - n_h} \qquad \text{and} \qquad \lim_{B \to 0} R_\text{H} = -\frac{n_e \mu_e^2 - n_h \mu_h^2}{(n_e \mu_e + n_h \mu_h)^2},$$

where $n_{e,h}$ and $\mu_{e,h}$ are the electron and hole carrier densities and mobilities of the two bands, respectively. In our measurements, we access a regime of high conductance and observe a nonlinearity of the Hall data around 5 T, as shown in Fig. S5a. From the corresponding derivative (Fig. S5b) it can be seen that at higher magnetic fields the Hall curve is saturated towards a linear behavior ($R_\mathrm{H}$ = constant), indicating that the high-field limit has been reached.

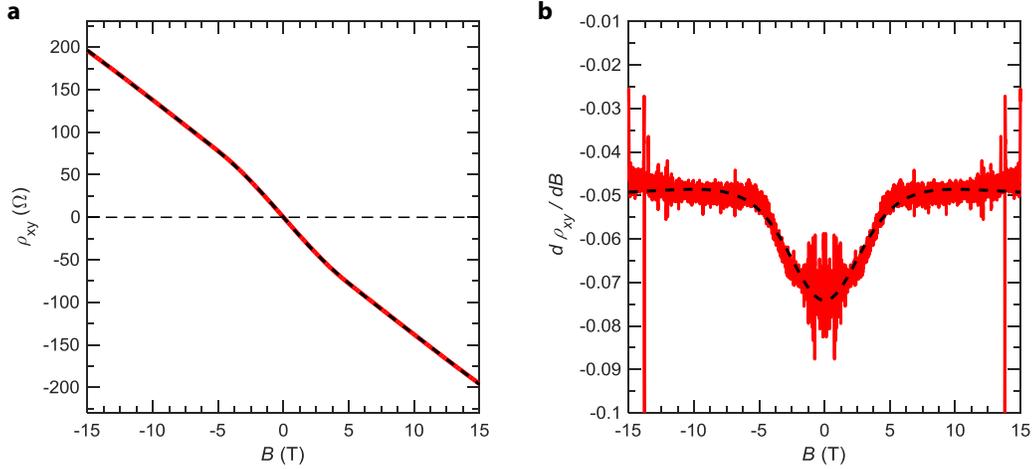

FIG. S5. (a) Experimental Hall curves for the highest electrostatic doping (red) and respective fit (dashed black). (b) Derivative of the curves presented in (a).

Since $R_\mathrm{H}(B \to \infty) < 0$, we can unequivocally ascertain that the higher-density band is electron-like. From the low-field behavior it can be readily seen that $R_\mathrm{H}(B \to \infty) > R_\mathrm{H}(B \to 0)$. To prove that this is only compatible with two electron-like bands, we will show analytically that our data cannot be described by an electron-hole scenario, irrespective of the mobility values. Since the higher-density band is electron-like, we consider the situation where the second band is hole-like, i.e., $n_e > n_h$, where $n_e = x n_h$ and $x > 1$. Then

$$\lim_{B \to \infty} R_\mathrm{H} = -\frac{1}{x n_h - n_h} = -\frac{1}{n_h} \frac{1}{(x-1)}$$

and

$$\lim_{B \to 0} R_\mathrm{H} = -\frac{x n_h \mu_e^2 - n_h \mu_h^2}{(x n_h \mu_e + n_h \mu_h)^2} = -\frac{1}{n_h} \frac{x \mu_e^2 - \mu_h^2}{(x \mu_e + \mu_h)^2}.$$

We now check whether this is compatible with our data, where $R_\mathrm{H}(B \to \infty) > R_\mathrm{H}(B \to 0)$.

$$R_\mathrm{H}(B \to \infty) > R_\mathrm{H}(B \to 0)$$
$$\frac{x \mu_e^2 - \mu_h^2}{(x \mu_e + \mu_h)^2} > \frac{1}{(x-1)}$$
$$\frac{x \mu_e^2 - \mu_h^2}{x^2 \mu_e^2 + 2 x \mu_e \mu_h + \mu_h^2} > \frac{1}{(x-1)}$$
$$(x-1)(x \mu_e^2 - \mu_h^2) > x^2 \mu_e^2 + 2 x \mu_e \mu_h + \mu_h^2$$
$$(\cancel{x^2}-x)\mu_e^2 - (x - \cancel{1})\mu_h^2 > \cancel{x^2 \mu_e^2} + 2 x \mu_e \mu_h + \cancel{+\mu_h^2}$$
$$-x(\mu_e^2 + \mu_h^2) > 2 x \mu_e \mu_h$$
$$-x > \frac{2 x \mu_e \mu_h}{\mu_e^2 + \mu_h^2}$$

The expression of the right hand side is always positive because $x$, $\mu_e$ and $\mu_h > 0$, and therefore the solution does not exist. This proves that our data cannot be described by an electron-hole scenario, irrespective of the mobility

values. Moreover, we show that the data is very well fitted by the two-band model with two electron-like bands (dashed black lines in Fig. S5a and b), which provides the carrier densities and mobilities shown in Fig. 2c of the main text.

---


\end{document}